\documentclass[prl,twocolumn,preprintnumbers,amsmath,amssymb,floatfix,superscriptaddress,showpacs]{revtex4}
\usepackage{graphicx}
\usepackage{graphics}
\usepackage{psfrag}
\usepackage{hyperref}
\usepackage{multirow}
\usepackage[sort&compress]{natbib}
\usepackage{color}
\usepackage{slashed}
\usepackage{verbatim}

\newcommand{\vare}{\varepsilon}

\newcommand{\rmi}{{\rm i}}

\begin{document}

\hypersetup{pdftitle={Equation of State of Ultracold Fermions in the 2D BEC-BCS Crossover Region}}
\title{Equation of State of Ultracold Fermions in the 2D BEC-BCS Crossover Region}
\date{\today}
\author{I. Boettcher}
\thanks{To whom correspondence should be addressed. E-mail: \href{mailto:I.Boettcher@thphys.uni-heidelberg.de}{I.Boettcher@thphys.uni-heidelberg.de}}
\affiliation{Institute for Theoretical Physics, Heidelberg University, D-69120 Heidelberg, Germany}
\author{L. Bayha}
\affiliation{Physikalisches Institut, Heidelberg University, D-69120 Heidelberg, Germany}
\author{D. Kedar}
\affiliation{Physikalisches Institut, Heidelberg University, D-69120 Heidelberg, Germany}
\author{P. A. Murthy}
\affiliation{Physikalisches Institut, Heidelberg University, D-69120 Heidelberg, Germany}
\author{M. Neidig}
\affiliation{Physikalisches Institut, Heidelberg University, D-69120 Heidelberg, Germany}
\author{M. G. Ries}
\affiliation{Physikalisches Institut, Heidelberg University, D-69120 Heidelberg, Germany}
\author{A. N. Wenz}
\affiliation{Physikalisches Institut, Heidelberg University, D-69120 Heidelberg, Germany}
\author{G. Z\"{u}rn}
\affiliation{Physikalisches Institut, Heidelberg University, D-69120 Heidelberg, Germany}
\author{S. Jochim}
\affiliation{Physikalisches Institut, Heidelberg University, D-69120 Heidelberg, Germany}
\author{T. Enss}
\affiliation{Institute for Theoretical Physics, Heidelberg University, D-69120 Heidelberg, Germany}

\begin{abstract}
We report the experimental measurement of the equation of state of a two-dimensional Fermi gas with attractive s-wave interactions throughout the crossover from a weakly coupled Fermi gas to a Bose gas of tightly bound dimers as the interaction strength is varied. We demonstrate that interactions lead to a renormalization of the density of the Fermi gas by several orders of magnitude. We compare our data near the ground state and at finite temperature to predictions for both fermions and bosons from Quantum Monte Carlo simulations and Luttinger-Ward theory. Our results serve as input for investigations of close-to-equilibrium dynamics and transport in the two-dimensional system.
\end{abstract}

\pacs{05.30.Fk, 67.85.Lm, 05.70.Ce, 03.75.Hh}

\maketitle

%\tableofcontents

The rich phenomenology of fermionic many-body systems reveals itself on very different scales of energy, ranging from solid state materials and ultracold quantum gases to heavy-ion collisions and neutron stars. Understanding the underlying mechanisms promises substantial advances both on a fundamental and technological level. Ultracold quantum gases provide a platform for the exploration of the macroscopic phases and thermodynamic properties of fermionic many-body Hamiltonians in a highly controlled manner \cite{RevModPhys.80.885}. In particular, using strongly anisotropic traps, it is possible to enter the 2D regime \cite{PhysRevLett.105.030404,PhysRevLett.106.105304,KoehlNature,PhysRevLett.108.045302,PhysRevLett.112.045301,PhysRevLett.114.230401} which is of large interest to the condensed matter community \cite{Loktev20011,Lee2006}.

The thermodynamic properties of a many-body system are encapsulated 
in its equation of state (EOS) $n(\mu,T, \{g_i\})$, which expresses the density $n$ as a function of chemical potential $\mu$, temperature $T$, and further
system parameters $\{g_i\}$ characterizing, for instance, the interactions between particles. For ultracold atoms with short-range attraction, the only additional parameter is the s-wave scattering length $a$. This universality allows one to describe different atomic species by the same EOS $n(\mu,T,a)$. The equilibrium EOS is also the basis for studying dynamics close to thermal equilibrium.

In this Letter, we report the experimental determination of the EOS of two-component fermions with attractive short-range interactions in the 2D BEC-BCS crossover regime. We tune the interaction strength using a Feshbach resonance to connect the well-known limits of a weakly attractive Fermi gas and a Bose gas of tightly bound dimers. We report the measurement of the finite temperature EOS in the intermediate, strongly correlated region and compare with theoretical predictions.

Our experimental setup consists of a population-balanced mixture of $N\sim100,000$ $^6$Li-atoms in the lowest two hyperfine states, which we denote by $|1\rangle$ and $|2\rangle$. The interactions between both species can be tuned by means of a magnetic Feshbach resonance \cite{Ketterle2008,PhysRevLett.110.135301}.  The atoms are trapped in a highly anisotropic trapping potential, which is radially symmetric to a high degree in the $xy$-plane and provides a tight confinement along the $z$-direction
with the aspect ratio of frequencies $\omega_x : \omega_y : \omega_z\approx 1:1:310$. A detailed description of the experiment is given in \cite{PhysRevLett.114.230401}. This strong anisotropy induces a quantum confinement of the many-body system, with discrete excitation spectrum in the $z$ direction and an effective 2D continuum of states in the $xy$ plane. A 2D system is realized if the interacting system is in its ground state  in the $z$-direction. Although the explicit form of this ground state is not known and depends on the 3D scattering length $a_{\rm 3D}$ \cite{2014arXiv1411.4703D}, the ratio $N(\omega_x/\omega_z)^2\sim 1$ close to unity 
indicates that the first excited state in $z$ direction will be partially populated.
In this quasi-2D regime, interactions can still be described by an effective 2D scattering length $a_{\rm 2D}$ \cite{PhysRevA.64.012706,PhysRevLett.112.045301,Levinsen2015}.

In order to parametrize the local density of the gas, we introduce the Fermi wave vector $k_{\rm F}=(2\pi n)^{1/2}$, where $n=n_1+n_2$ is the planar density and $n_\sigma=n/2$ is the density of atoms in state $|\sigma\rangle$. We further define the Fermi energy $\vare_{\rm F}=\hbar^2k_{\rm F}^2/(2M)$ and Fermi temperature $T_{\rm F}=\vare_{\rm F}/k_{\rm B}$, where $\hbar$ is Planck's constant, $k_{\rm B}$ Boltzmann's constant, and $M$ the atomic mass of $^6$Li. The 2D crossover \cite{PhysRevB.12.125,Miyake01061983,PhysRevLett.62.981,PhysRevLett.63.445,PhysRevLett.65.1032,ANDP19925040105,Petrov2003,PhysRevLett.103.165301,PhysRevLett.112.135302,PhysRevLett.115.115301} is parametrized by $\ln(k_{\rm F}a_{\rm 2D})$. In the BEC limit $\ln(k_{\rm F}a_{\rm 2D})\ll-1$ the system can be described as a 2D Bose gas with effective coupling constant $\tilde{g} \approx -2\pi/\ln(k_{\rm F}a_{\rm 2D})$. In the BCS limit $\ln(k_Fa_\text{2D})\gg1$ the thermodynamic properties approach that of an ideal Fermi gas.
In analogy to the 3D BEC-BCS crossover \cite{Chin2004,Zwierlein2005,Zwierlein2006,Zwerger}, the 2D gas undergoes a finite temperature phase transition to a superfluid state for all values of the  parameter $(k_{\rm F}a_{\rm 3D})^{-1}$ \cite{PhysRevLett.114.230401}. The associated transition, however, is of BKT type and the superfluid phase exhibits quasi long-range order \cite{Berezinskii1971,Berezinskii1972,Kosterlitz1973,Hadzibabic2006,Murthy2014,PhysRevLett.115.010401}.

For a 2D ultracold quantum gas with short-range attraction, a two-body bound state with binding energy $\vare_{\rm B}=\hbar^2/(Ma_{\rm 2D}^2)>0$ exists for all values of the coupling strength. This has to be contrasted to the 3D case, where a bound state only exists on the Bose side of the resonance. In both cases, the chemical potential for a single fermion becomes negative in the Bose limit and approaches $\mu \approx -\vare_{\rm B}/2$. The crossover point between the Bose and Fermi sides can be defined by the zero crossing of $\mu$ \cite{PhysRevLett.111.265301}.
The chemical potential shifted by the bound-state energy, $\tilde\mu = \mu + \vare_{\rm B}/2$, is positive at high phase-space densities.  At zero temperature, $n>0$ is equivalent to $\tilde\mu>0$.

In the experimental realization of the quasi-2D gas in an anisotropic 3D trap, the interaction strength $a_{\rm 2D}$ depends on the typical momenta of scattering particles and thereby on the filling in the trap with axial frequency $\omega_z$. One can write $a_{\rm 2D}=a_{\rm 2D}^{(0)} e^{-\frac{1}{2}\Delta w(\tilde{\mu}/\hbar\omega_z)}$, where $a_{\rm 2D} ^{(0)}$ is the scattering length in the dilute limit and $\Delta w$ is a positive function which reduces $a_{\rm 2D}$ at finite density \cite{PhysRevA.64.012706,PhysRevLett.112.045301,SOM}.
The correction to $a_{\rm 2D}^{(0)}$ vanishes in the Bose limit where $\tilde{\mu}\to0$, and becomes strongest in the Fermi limit where $\tilde{\mu} \simeq \vare_{\rm F}$.
In our experiment we have $\hbar\omega_z/k_{\rm B} =265\text{nK}$, which has to be compared with typical values $\tilde{\mu}_0=(40\dots200)\text{nK}$ and $T=(60\dots25)\text{nK}$ when going from the Bose to the Fermi limit. As most particles are in the center of the cloud, we approximate $a_{\rm 2D}$ and $\vare_{\rm B}$ by their central values using $\Delta w(\tilde{\mu}_0/\hbar\omega_z)$, giving $\Delta w\approx 0.2, 0.9,$ and $1.4$ in the Bose, crossover, and Fermi regimes, respectively.

We extract the  EOS of the homogeneous gas from the trapped system by using the local density approximation (LDA) which assigns a local chemical potential $\mu(\vec{r})=\mu_0-V(\vec{r})$ to each point $\vec{r}$ in the trapping potential $V(\vec{r})$ \cite{LandauStat}. Since $V(\vec{r})$ is known to a high precision, the homogeneous density $n(\mu,T)$ can be deduced from the measured local in-situ density of the inhomogeneous system $n(\vec{r}) = n(\mu_0-V(\vec{r}),T)$ once $\mu_0$ and $T$ have been determined \cite{HoZhou}.
The extraction of the homogeneous EOS from the trapped gas has been applied to both bosonic and fermionic systems and successfully compared with theoretical calculations \cite{Horikoshi22012010,Salomon,Navon07052010,Hung2011,PhysRevLett.107.130401,Ku03022012,PhysRevLett.113.020404,Vale20165}.

\textit{Low-temperature EOS.}---In order to  determine the low-temperature equation of state $n(\mu,T\to0,a_{\rm 2D})$ we extract $\tilde{\mu}_0$ from a Thomas--Fermi (TF) fit of the central region of the cloud. The TF model assumes locally $\vare_{\rm F} = c\cdot\tilde{\mu}$ for the central density region. This scaling is valid for large phase space densities (PSD) $n\lambda_T^2$, where $\lambda_T=(2\pi\hbar^2/(Mk_{\rm B}T))^{1/2}$ is the thermal wavelength of atoms. We find that the prefactor $c$ only weakly depends on the temperature and fitting range at sufficiently low temperature and high densities, which confirms the validity of the linear relation $\vare_{\rm F} \propto \tilde{\mu}$ (see the supplemental material (SM) \cite{SOM} for details). We fit $c$ in the intervals $I_A=[0.4,0.8]n_{\rm peak}$ and $I_B=[0.5,1]n_{\rm peak}$ for peak density  $n_{\rm peak}$, and define  $\tilde{\mu}_0=(\tilde{\mu}_{0,A}+\tilde{\mu}_{0,B})/2$ as the average value of both outcomes.

\begin{figure}[t]
\centering
\includegraphics[width=8.5cm]{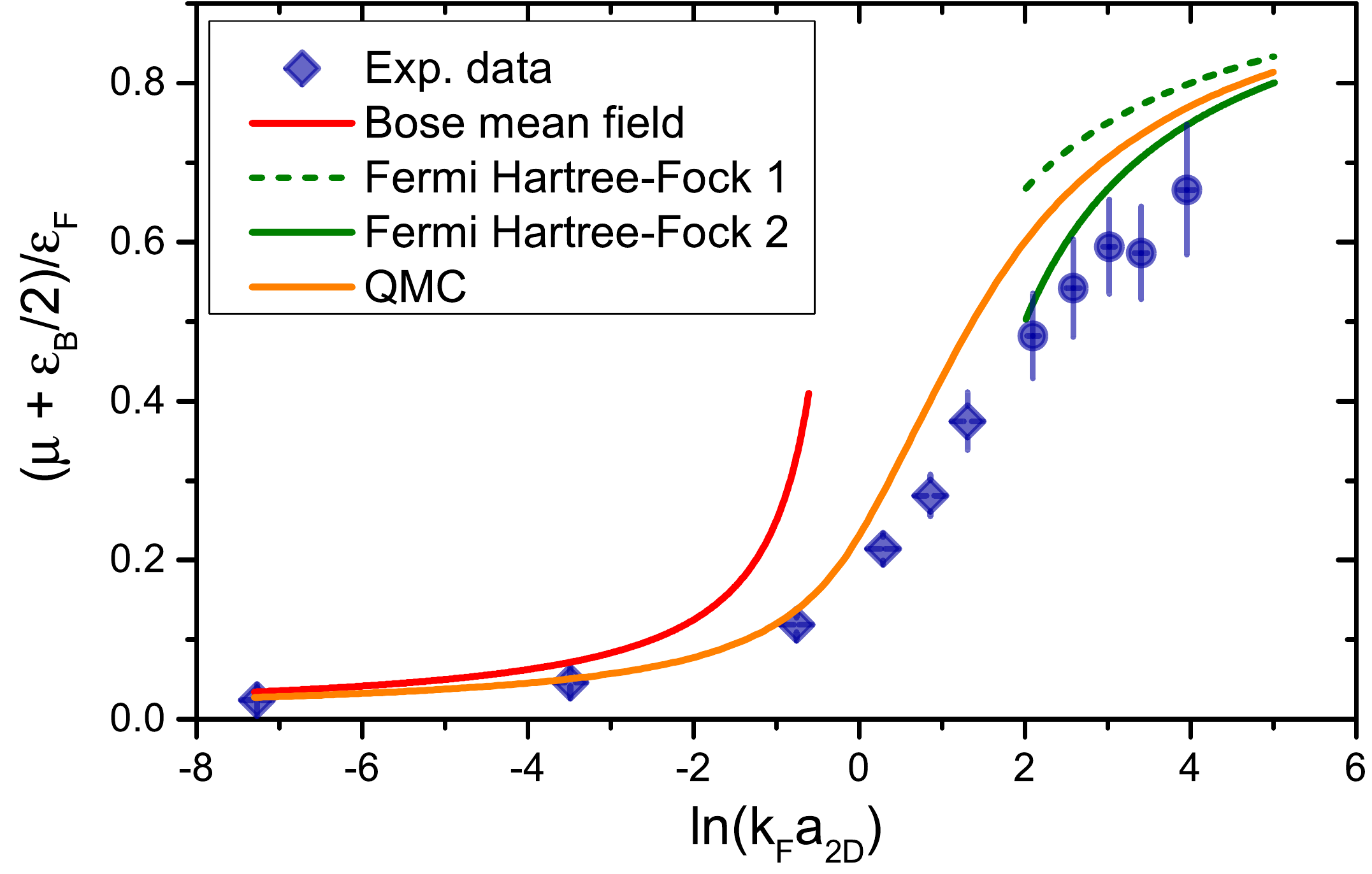}
\caption{Low-temperature EOS across the 2D BEC-BCS crossover. The experimental results are obtained from measurements of the quasi-2D gas at the lowest attainable temperatures, which corresponds to $T/T_{\rm F}\approx 0.05$ and $0.1$ on the Bose and Fermi sides. The data points shown as diamonds (circles) correspond to measurements in the superfluid (normal) phase. 
The solid red line on the Bose side corresponds to the mean field formula $\tilde{\mu}/\vare_{\rm F}=-\eta^{-1}/4$ with $\eta=\ln(k_{\rm F}a_{\rm 2D})$, whereas the dashed and solid green lines on the Fermi side display the non-selfconsistent and selfconsistent Hartree-Fock predictions $1/(1+\eta^{-1})$ and $1-\eta^{-1}$ for weakly attractive fermions. The orange line is the prediction for the ground state EOS from recent QMC calculations \cite{PhysRevA.92.033603}.}
\label{FigEOST0}
\end{figure}

In Fig.~\ref{FigEOST0} we show the low-temperature EOS across the 2D BEC-BCS crossover in terms of $\tilde{\mu}/\vare_{\rm F}=1/c$ vs. $\ln(k_{\rm F}a_{\rm 2D})$, where $k_{\rm F}$ corresponds to the peak density. The corresponding temperatures (60--25nK from left to right) are very low compared to $T_{\rm F}$ (1500--300nK). For the plot we averaged the TF slope
$c$ over 30 images at the lowest temperatures for each value of the interaction strength. We find $c$ to be weakly dependent on temperature as we increase $T/T_{\rm F}$ by $(40-100)\%$, which is a necessary condition for the applicability of the linear fit of the central region.  The statistical error of $\tilde{\mu}/\vare_{\rm F}$ is $10\%$ within the whole crossover. We estimate the error due to systematic uncertainties 
resulting from the absorption imaging, atoms in noncentral pancakes \footnote{We estimate that the presence of atoms in noncentral pancakes, i.e., neighboring sites of the standing wave potential, results in an overestimate of the planar density by at most 10\% \cite{SOM}.}, magnification, and the determination of the binding energy 
to be $15\%$ and $13\%$ on the Bose and Fermi sides, respectively \cite{PhysRevLett.114.230401,SOM}.

Our measured low-temperature equation of state connects both perturbative limits of the crossover.
In the Bose limit we compare our results with predictions for bosonic dimers of mass $2M$, dimer density $n_{\rm d}=n/2$, chemical potential $\mu_{\rm d}=2\tilde{\mu}$, and thermal wavelength $\lambda_{\rm d}=\lambda_T/\sqrt{2}$.
The interactions between dimers can be modelled by an effective 2D coupling strength $\tilde{g}= \sqrt{16\pi} \frac{0.6a_{\rm 3D}}{\ell_z}$  with $\ell_z=\sqrt{\hbar/(M\omega_z)}=0.551\mu\text{m}$ the oscillator length of axial confinement  \cite{Petrov2004}. 
We find $\tilde{\mu}/\vare_{\rm F}=0.024(2), 0.046(4), 0.12(1)$ for the data points with $\ln(k_{\rm F}a_{\rm 2D})\leq-0.71$ corresponding to effective boson coupling strengths $\tilde{g}=0.60, 1.07, 2.75$, respectively. This is in excellent agreement with the perturbative Bose gas formula $\tilde{\mu}/\vare_{\rm F}= \tilde{g}/(8\pi) = 0.024, 0.043, 0.11$. Furthermore, we verify $\tilde{g}\approx -2\pi/\ln(k_{\rm F}a_{\rm 2D})$ for very small $\tilde{g}$, which is a result of $a_{\rm 2D}\simeq a_{\rm 2D}^{(0)}$ on the Bose side, where the filling correction $\Delta w$ is small.

Far on the Fermi side, our low-temperature data is consistently below the Hartree-Fock (HF) prediction $\tilde{\mu}/\vare_{\rm F}\simeq\mu/\vare_{\rm F}=1-(\ln(k_{\rm F}a_{\rm 2D}))^{-1}$. However, the additional error due to systematic uncertainties \cite{SOM} is of the same size as the statistical one displayed in Fig. \ref{FigEOST0},  such that we find consistency with HF theory within the errors of our measurements.
Note that an extension of the BCS mean-field theory toward the crossover, which works reasonably well in 3D \cite{PhysRev.186.456,leggett80}, would give $\tilde{\mu}/\vare_{\rm F}=1$ for all interaction strengths in the 2D case and thus clearly misses the crossover physics \cite{PhysRevLett.114.110403}.  The ground-state equation of state has been investigated theoretically in \cite{PhysRevLett.106.110403,PhysRevA.91.011604,PhysRevA.92.023620,PhysRevA.92.033603}. In Fig.~\ref{FigEOST0} we compare our measurements to recent QMC simulations of the ground state \cite{PhysRevA.92.033603}\footnote{For the plot we use the piecewise defined interpolation function $f(x)$ with $x=\ln(2e^{-\gamma}k_{\rm F}a_{\rm 2D})$ given in Ref.~\cite{PhysRevA.92.033603}. From $\mu=\mbox{d}\epsilon/\mbox{d}n$ we then find $\tilde{\mu}/\vare_{\rm F}=1+f(x)+\frac{1}{4}f'(x)$.}. For a comparison of different theoretical predictions of $\tilde{\mu}/\vare_{\rm F}$ at zero temperature see, for instance, Ref. \cite{PhysRevA.92.023620}. Our data at small but finite temperature lie consistently below the zero-temperature prediction.

\textit{Finite-temperature EOS.}---While the temperature is constant within each atom cloud, it varies for every individual realization of the gas (``shot''). In our analysis we therefore determine $T$ and $\mu_0$ from each density profile and construct the dimensionless PSD $f_i(x,y) = f(\beta\tilde{\mu},\beta\vare_{\rm B}) = n\lambda_T^2$ and normalized density $h_i(x,y) = h(\beta\mu,\beta\vare_{\rm B}) = n/n_0$ for every shot $i$. Here, $n_0(\mu,T)=2\lambda_T^{-2}\ln(1+e^{\beta\mu})$ is the EOS of an ideal Fermi gas. Finally, we average $f_i$ and $h_i$ over 30--150 shots to obtain the EOS with very small statistical error, even though the thermodynamic parameters vary from shot to shot.
The values of  $T$ and $\mu_0$ can be found by different methods: whereas the TF fit determines $\mu_0$ from the dense central region of the cloud, fitting a reference EOS to the outer low-density regions gives both $T$ and $\mu_0$.

We first summarize the reference EOSs used in this work. In the perturbative Bose limit of small $\tilde{g}$, the outer wings are described by the HF formula $n_{\rm d}\lambda_{\rm d}^2 =-\ln(1-e^{\beta\mu_{\rm d}-(\tilde{g}/\pi)n_{\rm d}\lambda_{\rm d}^2})$. The Boltzmann limit for a gas of dimers or atoms, respectively, reads $n_{\rm d}=\lambda_{\rm d}^{-2} e^{\beta \mu_{\rm d}}=2\lambda_T^{-2}e^{2\beta\tilde{\mu}}$ and $n_\sigma=\lambda_T^{-2}e^{\beta\mu}$. The latter two formulas are elegantly connected by the second order virial expansion $n_\sigma \lambda_T^2=\ln(1+e^{\beta\mu})+2b_2e^{2\beta\mu}$ 
\cite{PhysRevA.88.043636,PhysRevLett.111.265301,SOM}. In the weakly interacting Fermi limit $b_2\to 0$, and $n_\sigma \lambda_T^{2}$ approaches the EOS of an ideal Fermi gas. On the Bose side, instead, $\vare_{\rm B}$ becomes large and the fermion fugacity $e^{\beta\mu}=e^{\beta(\tilde{\mu}-\vare_{\rm_B}/2)}$ is extremely small, suppressing the first term of the EOS. However, 
$b_2=e^{\beta\vare_{\rm B}}$ up to exponentially small corrections and we recover the bosonic Boltzmann EOS $n_\sigma = 2 \lambda_T^{-2} e^{\beta(2\mu+\vare_{\rm B})} = \lambda_{\rm d}^{-2} e^{\beta\mu_{\rm d}}$. Hence, the second order virial expansion has the correct limiting behavior and provides a well-defined reference EOS throughout the crossover.

We apply the HF formula for the perturbative Bose gas to determine $T$ and $\mu_0$ only for $B=692\text{G}$ where $\tilde{g}=0.60$. For the remaining magnetic fields, $B[\text{G}]=732-922$, the central chemical potential $\tilde{\mu}_0$ is determined from the TF fit of the central region. 
The temperature is estimated by $T=(T_{\rm V}+T_{\rm B})/2$, where $T_{\rm V}$ and $T_{\rm B}$ are obtained from second order virial and Boltzmann fits to the outer region, respectively. This choice is motivated by the observation that $T_{\rm V}$ and $T_{\rm B}$  
give upper and lower bounds on the true temperature
for the interaction strengths considered here. We quantitatively compare different methods to obtain $T$ and $\mu_0$ in the SM \cite{SOM}. In particular, we show that $\mu_0$ obtained from the virial and Boltzmann fits agrees well with the one from the TF fit, which also supports the validity of the TF assumption.

\begin{figure}[t]
\centering
\includegraphics[width=8.5cm]{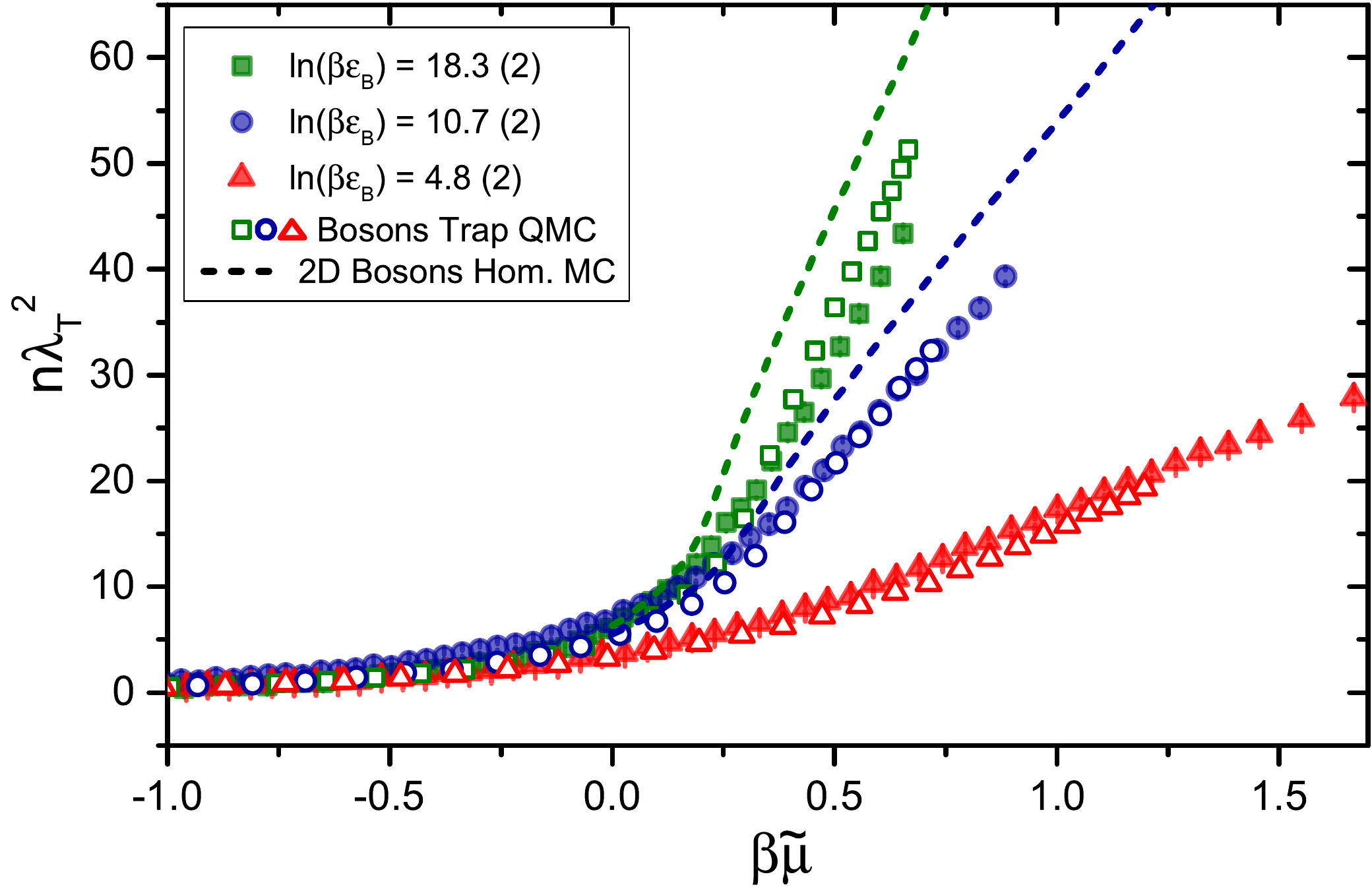}
\caption{Phase space density on the Bose side of the crossover. The experimental data points are shown as filled shapes. We compare to bosonic theory with effective coupling strengths $\tilde{g}=0.60, 1.07, 2.75$. Open shapes represent the EOS extracted from the QMC simulation of the quasi-2D Bose gas \emph{trapped} in an external potential with similar parameters as employed in experiment. The dashed curves show the classical MC prediction for the weakly coupled \emph{homogeneous} 2D Bose gas from Ref.\ \cite{PhysRevA.66.043608} extrapolated to large values of $\tilde{g}$. For moderate densities we find good agreement between all three approaches. For large densities, however, experiment and trapped QMC deviate from the classical homogeneous result which scales like the mean field prediction $n_{\rm d}\lambda_{\rm d}^2 = \frac{2\pi}{\tilde{g}} \beta\mu_{\rm d}+\ln(\frac{2\tilde{g}}{\pi}n_{\rm d}\lambda_{\rm d}^2-2\beta\mu_{\rm d})$. This may be due to quantum effects appearing at large $\tilde{g}$, as well as the axial confinement.}
\label{FigPSDbose}
\end{figure}

\begin{figure*}[t]
\centering
\includegraphics[width=17cm]{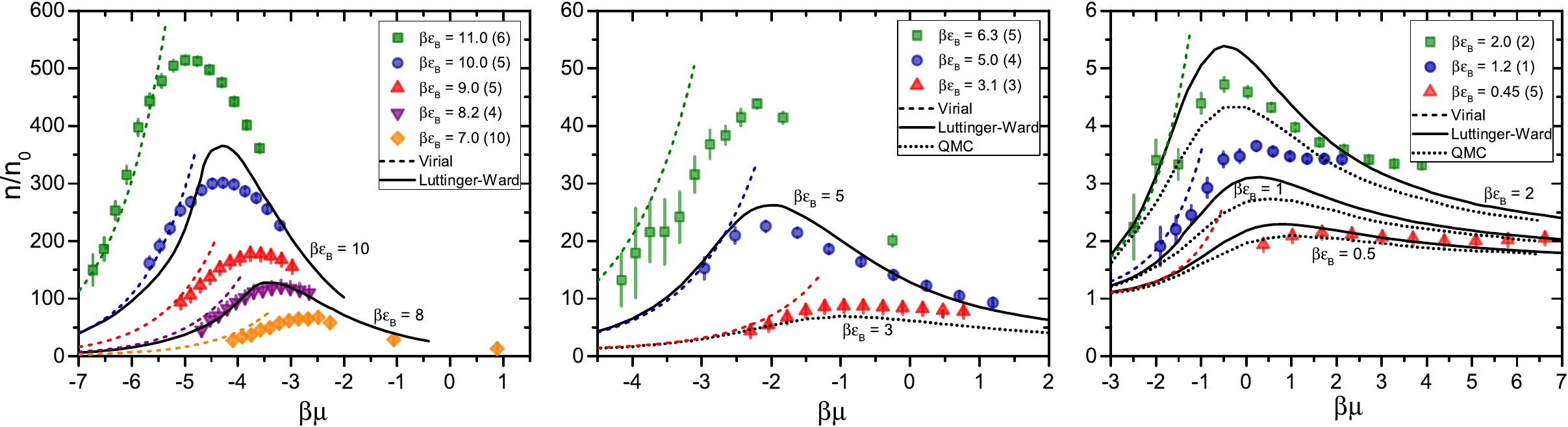}
\caption{EOS in the crossover regime shown as the density normalized by the ideal Fermi gas $n_0(\mu,T)=2\lambda_T^{-2}\ln(1+e^{\beta\mu})$. The experimental data points (filled shapes) are compared with the second order virial expansion at low values of $\beta\mu$ (coloured dashed lines). The displayed errors are purely statistical, with systematic uncertainties estimated at 13--15$\%$ and given for each data set in Table IV in the SM \cite{SOM}. We compare our results with theoretical predictions in the 2D BEC-BCS crossover from Luttinger-Ward theory (\cite{PhysRevLett.112.135302}, solid black lines) and fermionic QMC simulations (\cite{PhysRevLett.115.115301}, dotted black lines), with the corresponding values of $\beta\vare_{\rm B}$ labelling each curve. Note that the vertical scale differs by a factor of 10 in each panel.}
\label{FigPSDfermi}
\end{figure*}

\textit{Bose EOS.}---Figure \ref{FigPSDbose} shows the EOS on the Bose side of the crossover. 
Due to the exponentially large binding energy in the Bose limit, the logarithmic dependence of the EOS on $\beta\vare_{\rm B}$ in $n\lambda_T^2=f(\beta\tilde{\mu},\beta\vare_{\rm B})$ can be replaced by the $\tilde{g}$ dependence in $n\lambda_T^2=F(\beta\tilde{\mu},\tilde{g})$, where $F(x,\tilde{g})$ can directly be compared with
bosonic theory for coupling $\tilde{g}$. The experimental data correspond to $\tilde{g}=0.60, 1.07, 2.75$. The plotted curves represent the average $F(x,\tilde{g})$ from approximately 30 individual shots for each interaction strength.

We compare our results on the Bose side to classical Monte Carlo (MC) and Quantum Monte Carlo (QMC) simulations of bosons in order to understand whether the fermionic system can be described purely in terms of bosons. The classical MC computations are valid for the weakly coupled, homogeneous 2D Bose gas in the fluctuating regime \cite{Prokofev2001,PhysRevA.66.043608}. In our case $\tilde{g}$ is large, and quantum effects are expected to modify the result. Even so, the critical temperature for $\tilde{g}<3$ is well described by extrapolating MC to large $\tilde{g}$ \cite{PhysRevLett.110.145302,PhysRevLett.114.255302,PhysRevLett.115.010401}. We also analyze density profiles obtained from QMC simulations of a trapped Bose gas with similar trapping parameters as in the experiment; see Refs.~\cite{Holzmann2008,Holzmann2010,PhysRevLett.115.010401} for details. We extract the temperature and chemical potential of the QMC density profiles from the low-density Boltzmann regime and apply LDA to obtain the EOS $F(\beta\tilde{\mu},\tilde{g})$.

We find good agreement of our measurements with the EOS extracted from the QMC profiles.
The deviations between both EOSs are within the systematic errors in the determination of $n$, $T$ and $\mu_0$ and we conclude that the fermionic system with $\tilde{g}\leq 2.75$ is well-described by a strongly coupled quantum gas of dimers. 
Both experiment and QMC are, however, well below the classical MC predictions for large $\beta\tilde{\mu}$
and mean field theory. There are two effects which could explain this behavior: On the one hand, quantum fluctuations become important for large $\tilde{g}$ and high densities. On the other hand, both experiment and QMC are performed in a quasi-2D setting with nonzero extent in the $z$-direction.

\textit{Crossover EOS.}---Figure~\ref{FigPSDfermi} shows the EOS in the strongly correlated crossover regime between the Bose and Fermi limits. The EOS $h(x,y)=n/n_0$ is sampled over approximately 150 shots for each of the magnetic fields $B[\text{G}]=812, 832, 852, 892$. 
We compare our results with theoretical predictions for the homogeneous 2D BEC-BCS crossover from Luttinger--Ward (LW) theory \cite{PhysRevLett.112.135302} and fermionic QMC simulations \cite{PhysRevLett.115.115301}. 
Our comparison covers a substantial renormalization by two orders of magnitude in the density $n/n_0$. It reveals a maximum in $n/n_0$ characteristic of the density driven crossover in 2D \cite{PhysRevLett.112.135302}. We find that the maximum of height $2 e^{\beta\vare_{\rm B}/2}$ is reached at $\beta\mu \simeq -\beta \vare_{\rm B}/2+\ln(2)$ for large $\beta\vare_{\rm B}$.
The origin of this scaling can be understood from the virial expansion in the Bose limit: $n_\sigma\lambda_T^2 \approx 2\exp(2\beta\tilde\mu)= 2$ at $\beta\tilde\mu=0$, which implies $n/n_0 \approx 2/\ln(1+e^{-\beta\vare_{\rm B}/2}) \approx 2 e^{\beta\vare_{\rm B}/2}$ at $\mu=-\vare_{\rm B}/2$; see the SM \cite{SOM} for details. The difference between the LW and QMC EOSs lies within our systematic errors from the $T$ and $\mu_0$ determination and thus cannot be resolved with the present analysis. 
In the Fermi limit where $\tilde\mu_0/\hbar\omega_z\lesssim 0.75$ is largest, we observe that the filling correction $\Delta w$ shifts the EOS slightly upward. This effect is minimized for small particle numbers.
In a recent work by Fenech et al.\ \cite{Vale20165} the EOS in the normal phase for $\beta\vare_{\rm B}<0.5$ has been determined using $^6$Li atoms in the 2D regime.

In this work we have measured the EOS of ultracold fermions in the BEC-BCS crossover in a strongly anisotropic confinement. Our results connect the perturbative Bose gas, the strongly interacting Bose gas, the strongly interacting fermionic superfluid in the crossover regime, and the perturbative Fermi liquid as we tune the effective 2D scattering length $a_{\rm 2D}$ using a Feshbach resonance. Our EOS data covers both the low and intermediate temperature thermodynamics of the system. We compare with bosonic and fermionic quantum many-body theory and find a remarkably strong renormalization of the density $n/n_0$. These results provide a basis for phenomenological computations such as hydrodynamic models of the cloud.

\acknowledgments We gratefully acknowledge inspiring discussion with
M. Bauer, M. Holzmann, T. Lompe, and J. M. Pawlowski, and thank J. Drut and H. Shi
for sharing QMC data. This work has been supported by the ERC starting grant 279697,
the ERC advanced grant 290623, the Helmholtz Alliance HA216/EMMI, and the Heidelberg Center for Quantum Dynamics.

%%%%%%%%%%   REFERENCES   %%%%%%%%%%

\bibliographystyle{apsrev4-1}
\bibliography{refs_2d_eos_exp} %no .bib! %

\cleardoublepage

\section{Supplemental Material}

\section{Thomas--Fermi Conditions}\label{AppTF}
We obtain the low-temperature EOS and central chemical potential $\tilde{\mu}_0$ from a TF fit to the central region of the cloud. This method has the advantage of being independent of a particular model for the EOS. It applies when the result is independent of temperature and fitting range, as we discuss in this section.

We first motivate the linear scaling $\vare_{\rm F}(\mu) =c\ \tilde{\mu}$ at zero temperature. In particular, we show that the density does not scale like $\mu$, although $\mu$ is positive on the Fermi side of the crossover. Due to the nonvanishing binding energy $\vare_{\rm B}>0$ for all values of the interaction strength, the energy density (internal energy per volume) of the homogeneous system can be written as
\begin{align}
 \epsilon(n) = -n_{\rm d} \vare_{\rm B} + \delta\epsilon(n) = - n \frac{\vare_{\rm B}}{2} + \delta\epsilon(n),
\end{align}
where $n_{\rm d}=n/2$ is the density of pairs. The chemical potential is found from
\begin{align}
\mu(n) = \frac{\mbox{d}\epsilon(n)}{\mbox{d} n } = -\frac{\vare_{\rm B}}{2} + \frac{\mbox{d}\delta\epsilon(n)}{\mbox{d}n} =: -\frac{\vare_{\rm B}}{2} + \delta\mu(n).
\end{align}
The first term is independent of the density and thus does not contribute to the compressibility of the sample. This also uniquely defines the second term $\delta\mu(n)$. We write $\delta\mu(n)=\vare_{\rm F} /c$, where the dimensionless constant $c$ depends on the crossover parameter $\ln(k_{\rm F}a_{\rm 2D})$. Inverting the relation for $\vare_{\rm F}$ we arrive at
\begin{align}
 \vare_{\rm F} = c \Bigl(\mu+\frac{\vare_{\rm B}}{2}\Bigr) = c\ \tilde{\mu}.
\end{align}
This is the claimed scaling and we further conclude that $\delta\mu(n)=\tilde{\mu}(n)$. The quantity $\tilde{\mu}(n)$ thus encodes the difference between the interacting many-body system and a noninteracting gas of dimers.

To justify the TF model at nonzero temperature  we express the full EOS of the system in terms of the Fermi energy as
\begin{align}
 \label{tf1} \vare_{\rm F}(\mu,T,a_{\rm 2D}) = \tilde{\mu} \cdot g(\beta\tilde{\mu},\beta\vare_{\rm B})
\end{align}
with a dimensionless function $g(x,y)$. For large $\beta\tilde{\mu}$ (which corresponds to a large PSD) we can expand this expression according to
\begin{align}
 \label{tf2} \vare_{\rm F}(\mu,T,a_{\rm 2D}) = \tilde{\mu}\Bigl(g_0(\beta\vare_{\rm B}) + \frac{1}{\beta\tilde{\mu}} g_1(\beta\vare_{\rm B}) +\dots\Bigr)
\end{align}
with $g_0(y)=g(\infty,y)$ and $g_1(y)=\partial_x g(\infty,y)$. The leading term corresponds to the TF model with $c=g_0(\beta\vare_{\rm B})$.

\begin{figure}[t]
\centering
\includegraphics[width=7.5cm]{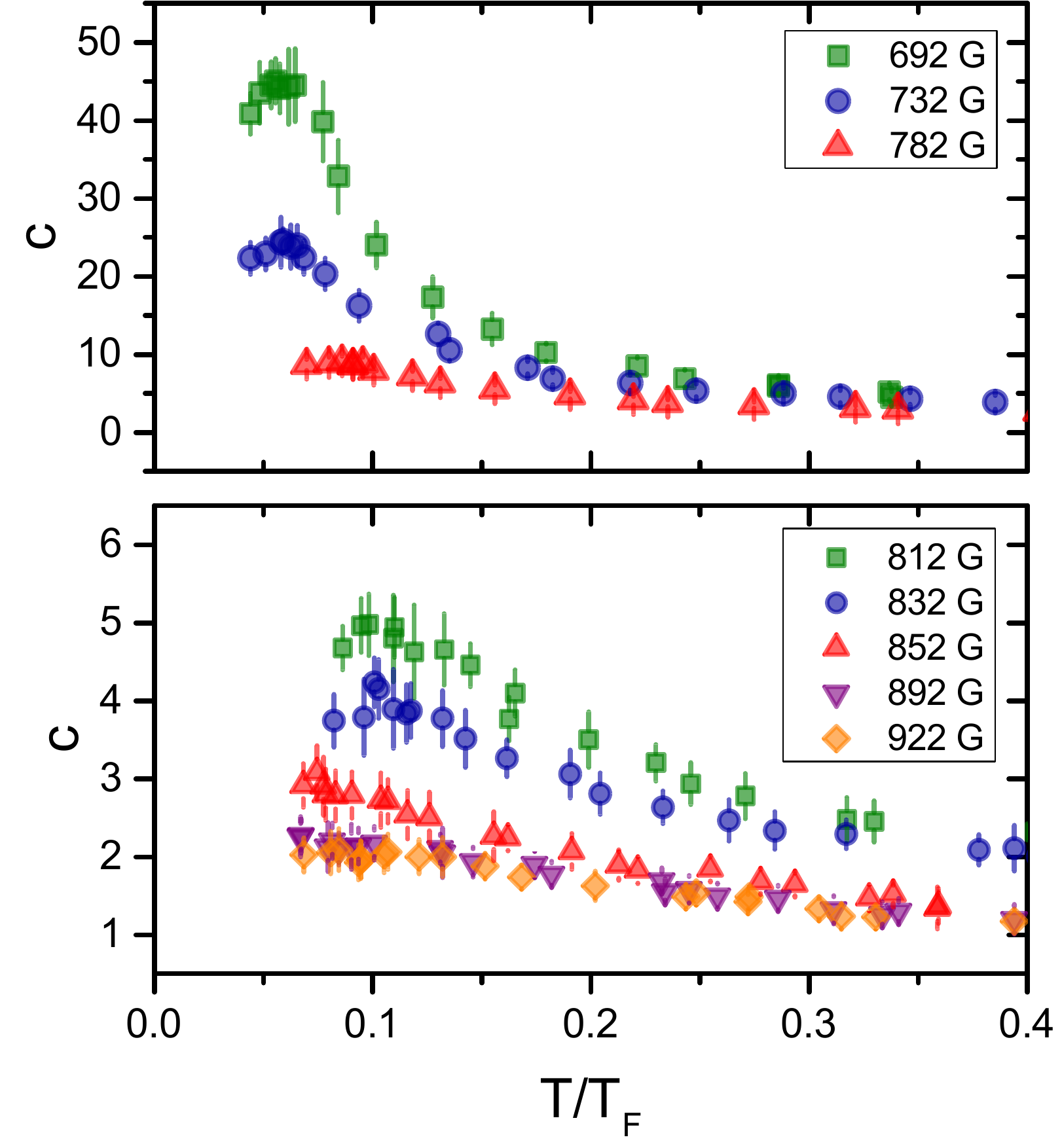}
\caption{Temperature dependence of the TF parameter $c$ on the Bose side (upper panel) and in the crossover regime (lower panel). We find $c$ to be approximately temperature independent for an increase of $T/T_{\rm F}$ by $40\%$ and $100\%$ on the Bose and Fermi side of the crossover.}
\label{Figc}
\end{figure}

Below we demonstrate that the dependence of $g(\beta\vare_{\rm B})$ on the binding energy is typically logarithmic in 2D. As a consequence, $c$ is to a good approximation temperature-independent for a fixed magnetic field $B$. We can employ this property to formulate a measure for the breakdown of the TF approximation: The validity of the expansion (\ref{tf2}) requires $c$ to be independent of temperature. In Fig.\ \ref{Figc} we show the results of the fit of $c$ in the Bose and crossover regime as a function of $T/T_{\rm F}$. Each data point corresponds to the average of approximately 30 shots. We observe $c$ to be approximately temperature independent for an increase of $T/T_{\rm F}$ by $40\%(100\%)$ on the Bose (Fermi) side.

To derive a second condition for the applicability of the TF fit, we note that a strictly linear dependence $\vare_{\rm F}\propto \tilde{\mu}$ would result in a constant $c$ which is independent of the chosen fitting range. If the function $\vare_{\rm F}(\tilde{\mu})$ is slightly nonlinear, the effective $c$ will vary with a change in the fitting range. To quantify this influence we fit our data in three ranges of densities: (I) $[0.4,0.8]n_{\rm Peak}$, (II) $[0.5,1]n_{\rm Peak}$, (III) $[0.7,1]n_{\rm Peak}$, where $n_{\rm Peak}$ is the peak density. For a function $\vare_{\rm F}(\tilde{\mu})$ which is slightly bending upwards for increasing $\tilde{\mu}$, the fitted value for $c$ would be smallest in (I) and largest in (III). We find the variation of $c$ with the fitting range to be compatible within the errors in all three intervals. We observe the deviations to be smaller on the Bose side (5-10$\%$) and to become larger as we approach the crossover regime and Fermi side (10-15$\%$).  There is another contribution to this effect from the influence of the third dimension. As shown below, there is a $\tilde{\mu}$-dependence of $\vare_{\rm B}$, which results in $\ln(\beta\vare_{\rm B}) = \ln(\beta\vare_{\rm B}^{(0)})+\Delta w (\frac{\tilde{\mu}}{\hbar\omega_z})$.  Still, we find that both $c_{\rm I}$ and $c_{\rm III}$ agree with $c_{\rm II}$ within the errors.

To obtain a reliable estimate for the true value we define the mean as
\begin{align}
 \label{tf3} C = \frac{C_{\rm I}+C_{\rm II}}{2},
\end{align}
where either $C=\tilde{\mu}_{0,\rm TF}, C=\tilde{\mu}/\vare_{\rm F}$, or $C=\ln(k_{\rm F}a_{\rm 2D})$, and $C_{\rm I/II}$ is the average of the observable over approximately 30 shots with the TF model applied to the fitting range I/II. The corresponding statistical error is given by $\Delta C = \sqrt{\frac{1}{2}(\Delta C_{\rm I}^2+\Delta C_{\rm II}^2)}\approx \Delta C_{\rm I/II}$. The fitting range dependence of the TF model results in a systematic error in the determination of the low-temperature EOS. The further error due to systematic uncertainties is discussed below.

Whereas we determine the low-temperature EOS from the measurements at the lowest attainable temperatures and estimate the error as outlined in the previous paragraph, it is important to estimate the temperature effect on the observable $1/c$. For this we quantify the constant behavior of the function $c(T/T_{\rm F})$ for small $T/T_{\rm F}$ (i.e. large phase space densities). We estimate the size of the plateau where $c$ is constant by determining the plateau's left and right boundaries $(T/T_{\rm F})_{\rm min} < (T/T_{\rm F})_{\rm max}$ from Fig. \ref{Figc}. 

For a fixed temperature $T$ we define $1/c$ as in Eq. (\ref{tf3}) according to $(1/c)_T=\frac{(1/c)_{\rm I}+(1/c)_{\rm II}}{2}$ to average the fitting range dependence and obtain the corresponding error $\Delta(1/c)_{T}=\sqrt{\frac{\Delta(1/c)_{\rm I}^2+\Delta(1/c)_{\rm II}^2}{2}}$. We then average over the data points for those temperatures $T_i$ within the plateau, $T_{\rm min}\leq T_i \leq T_{\rm mx}$. Let us assume that this comprises $N$ data points. We then define the temperature average as $\langle 1/c\rangle_{\rm temp} = \frac{1}{N}\sum_{i=1}^N (1/c)_{T_i}$. The observable $\langle1/c\rangle_{\rm temp}$ has two main error sources: The fitting range dependence, and the spread of the values of $(1/c)_T$ with $T$. We generalize the fitting range error according to $\Delta(1/c)_{\rm range}= \sqrt{\frac{1}{N}\sum_{i=1}^N(\Delta(1/c)_{T_i})^2}$ and define the error due to averaging over temperatures  as the standard deviation $\Delta(1/c)_{\rm avg}=\sqrt{\frac{1}{N}\sum_{i=1}^N[(1/c)_{T_i}-\langle1/c\rangle_{\rm temp}]^2}$. The total error of $\langle 1/c\rangle_{\rm temp}$ is then given by $\Delta(1/c)_{\rm temp}=\sqrt{\Delta(1/c)_{\rm range}^2+\Delta(1/c)_{\rm avg}^2}$. We find that $\Delta (1/c)_{\rm avg}$ is significantly smaller than $\Delta(1/c)_{\rm range}$ and $\Delta(1/c)_{\rm temp}\simeq \Delta(1/c)_{T_{\rm min}}$.

\begin{table}[t!]
\begin{tabular}{|c||c|c|c|}
 \hline $B[\text{G}]$ & $t=\frac{(T/T_{\rm F})_{\rm max}}{(T/T_{\rm F})_{\rm min}}-1$ & $(1/c)_{T_{\rm min}}$  & $\langle 1/c \rangle_{\rm temp}$ \\
 \hline 692 & $40\%$ & 0.024(2) & 0.024(2)\\
 \hline 732 & $49\%$ & 0.046(4) & 0.045(5)\\
 \hline 782 & $44\%$ & 0.12(1) & 0.12(1)\\
 \hline 812 & $38\%$ & 0.21(1) & 0.22(2)\\
 \hline 832 & $43\%$ & 0.28(3) & 0.27(3)\\
 \hline 852 & $52\%$ & 0.38(4) & 0.38(4)\\
 \hline 892 & $97\%$ & 0.48(5) & 0.49(6)\\
 \hline 922 & $93\%$ & 0.54(6) & 0.53(6) \\
 \hline 952 & $93\%$ & 0.59(6) & 0.58(7)\\
 \hline 982 & $96\%$ & 0.59(6) & 0.59(7)\\
 \hline 1042 & $106\%$ & 0.67(8) & 0.64(7)\\
 \hline
\end{tabular}
\caption{Thomas--Fermi fit. The slope $c$ is approximately constant for small values of $T/T_{\rm F}$ in the interval $(T/T_{\rm F})_{\rm min}\leq T \leq (T/T_{\rm F})_{\rm max}$, see Fig. \ref{Figc}. We show the size of this interval (plateau) in terms of the relative increase $t$ in $T/T_{\rm F}$. We estimate the variation within this temperature-plateau by comparing the low-temperature EOS $1/c=\tilde{\mu}/\vare_{\rm F}$ at the lowest attainable temperature $T_{\rm min}$ with the one averaged over the plateau. We label these $(1/c)_{T_{\rm min}}$ and $\langle 1/c\rangle_{\rm temp}$, respectively. We observe the result of the temperature averaging within the plateau to hardly modify the value of $1/c$ and its error. The observable $(1/c)_{T_{\rm min}}$ coincides with $1/c$ shown in Fig. 1 of the main text and displayed in Tab. \ref{Taba2D} below.}
\label{TabSomcTemp}
\end{table}

With the definitions of the previous paragraph we are able to compare the value of $(1/c)_{T_{\rm min}}$ at the lowest attainable temperature $T_{\rm min}$ and the average $\langle 1/c\rangle_{\rm temp}$ over the plateau $T_{\rm min}\leq T\leq T_{\rm max}$. We show the results of this comparison in Tab. \ref{TabSomcTemp}. The ratio $t= (T/T_{\rm F})_{\rm max}/ (T/T_{\rm F})_{\rm min}-1$ gives the relative increase in $T/T_{\rm F}$. The displayed values of $1/c$ thus provide a quantitative statement on the size of the plateau. We find that the central value $\langle 1/c\rangle_{\rm temp}$ is at most slightly shifted with respect to $(1/c)_{T_{\rm min}}$ and that $\Delta(1/c)_{\rm avg}$ is small such that $\Delta(1/c)_{\rm temp}\simeq\Delta(1/c)_{T_{\rm min}}$. This indicates that the spread in temperatures of $(1/c)_T$ is small. Furthermore, the error due to the fitting range dependence is also only slightly changed due to averaging over the temperature-interval $T_{\rm min}\leq T \leq T_{\rm max}$. We conclude that the low-temperature EOS is faithfully captured by using either $(1/c)_{T_{\rm min}}$ or $\langle 1/c\rangle_{\rm temp}$. For the remaining analysis we employ the former choice.

\section{Temperature determination}\label{AppTdet}
In order to determine the temperature of the sample we fit the outer region of the cloud to a reference EOS. Here we quantitatively compare different approaches and motivate the particular choices employed in the main text. 

As our strategy for obtaining the EOS relies on a shot-by-shot determination of $T$ and $\mu_0$, we cannot use the temperature determination from the tail of the momentum distribution from Ref.\ \cite{PhysRevLett.114.230401}, as it only gives the average over sets of shots. In fact, we find relatively large fluctuations from shot to shot for the temperature. Therefore, we display the median and the median deviation in the tables below.

We first discuss the fitting ranges applied for the temperature fits.
In each density profile we find that the local density fluctuations $\Delta n_\sigma \lesssim n_\sigma^\text{(th)}$ are below a threshold $n_\sigma^{(\rm th)}\approx0.02 \mu\text{m}^{-2}$ determined by the temperature of the gas as well as the imaging setup. The value of $n_\sigma^\text{(th)}$ sets the density scale where the temperature fit can be applied. Since the density fluctuations average to zero, $\langle \Delta n_\sigma\rangle=0$, the density profile in the wings can safely be fitted also below the threshold, $n_\sigma < n_\sigma^\text{(th)}$.
With temperatures $T\simeq60-25\text{nK}$ on the Bose and Fermi sides we find $\lambda_T^2n_\sigma^{(\rm th)} \simeq 0.2-0.4$.

To test for the fitting range dependence of the temperature fits of $n(\mu,T)$ we apply each reference EOS discussed below to two fit regimes: These are (A) the density range from $0.1n_{\rm peak}$ to zero and (B) the $\mu$-interval $\Delta\mu=\pm 20\text{nK}$ centered at $n_\sigma^{(\rm th)}$. We further vary the size of $\mu$-bins in the range $1-5\text{nK}$. In all cases we obtain results for $T$ and $\mu_0$ in close agreement with each other. For the following we thus restrict the discussion to fitting range (A) and a bin size of approximately $1\text{nK}$.

On the Bose side of the crossover, i.e., for magnetic fields $B[G]=692, 732, 782$, the low density region for small $\tilde{g}$ is given by the  HF formula 
\begin{align}
 \label{temp1} \text{HF:}\ n_{\rm d}\lambda_{\rm d}^2 =-\ln(1-e^{\beta\mu_{\rm d}-(\tilde{g}/\pi)n_{\rm d}\lambda_{\rm d}^2}).
\end{align}
For 692G we have $\tilde{g}=0.60$ and HF theory should work reasonably well as has been demonstrated for bosons with $\tilde{g}\leq 0.5$ \cite{Hung2011}. For larger couplings the HF approximation breaks down. A simpler formula for the Bose limit is provided by a bosonic Boltzmann formula $n_{\rm d}\lambda_{\rm d}^2 = e^{\beta\mu_{\rm d}}$. In the QMC calculations of the Bose gas we find the density wings of the interacting Bose gas at 782G ($\tilde{g}=2.75$) to  be close to the bosonic Boltzmann formula even for moderate densities where deviations would be expected. The Boltzmann EOS is thus a trustworthy reference for the wings in this case. For the intermediate interaction strength $\tilde{g}=1.07$ at a magnetic field of 732G, both HF and bosonic Boltzmann formula give the same temperature, so we can use the Boltzmann result.

A continuous interpolation between the bosonic and fermionic limits is provided by the second order virial expansion
\begin{align}
 \label{temp2} \text{V:}\ n_\sigma \lambda_T^2 = \ln(1+e^{\beta\mu}) + 2b_2 e^{2\beta\mu},
\end{align}
where 
\begin{align}
 \label{temp3} b_2 = e^{\beta\vare_{\rm B}} -\int_{-\infty}^\infty \mbox{d}s \ \frac{\exp\{{-e^{s}/(2\pi)}\}}{\pi^2+(s-\ln(2\pi\beta\vare_{\rm B}))^2}
\end{align}
is the (interaction induced correction to the) second virial coefficient \cite{PhysRevA.88.043636,PhysRevLett.111.265301}. Equation (\ref{temp2}) is exact for small fermion fugacity $z=e^{\beta\mu}$ in the whole 2D BEC-BCS crossover. Its applicability for moderate values of $z$, however, is not guaranteed. For large $\beta\vare_{\rm B}$, $z\to 0$ and $b_2\simeq e^{\beta\vare_{\rm B}}$, such that Eq. (\ref{temp2}) approaches the bosonic Boltzmann formula. As $\beta\vare_{\rm B}$ becomes smaller, the second order virial expansion approaches the ideal Fermi gas formula. A comparison to LW and QMC theory, both of which should be reliable for sufficiently small $\beta\vare_{\rm B}$, indicates that the second order virial expansion is likely to overestimate the temperature of the sample, resulting in a curve $n/n_0$ vs. $\beta\mu$ which is too steep for small $\beta\mu$.

In order to estimate how much the second order virial expansion deviates from the true EOS we define the effective temperature exponent
\begin{align}
 \label{temp4} \alpha_{\rm eff}(\mu,T) = \frac{\mbox{d}\ln(n_\sigma\lambda_T^2)}{\mbox{d}\ln z} = \frac{1}{n_\sigma \lambda_T^2}\frac{\mbox{d}(n_\sigma\lambda_T^2)}{\mbox{d}(\beta\mu)}.
\end{align}
For an EOS given by $n_\sigma\lambda_T^2 = z^\alpha$ we have $\alpha_{\rm eff}=\alpha$ for all $\mu$ and $T$. For instance, the fermion Boltzmann gas corresponds to $\alpha_{\rm eff}=1$. In contrast, for large $b_2$, the second order virial expansion gives $\alpha_{\rm eff}=2$, in accordance with a bosonic Boltzmann gas. In general, $\alpha(\mu,T)$ will not be constant. Here we are interested in its value for $\beta\tilde{\mu}\in \mathcal{I}=[-2,0]$, which is the typical fitting range for the temperatures in our analysis.

\begin{figure}[t]
\centering
\includegraphics[width=8.5cm]{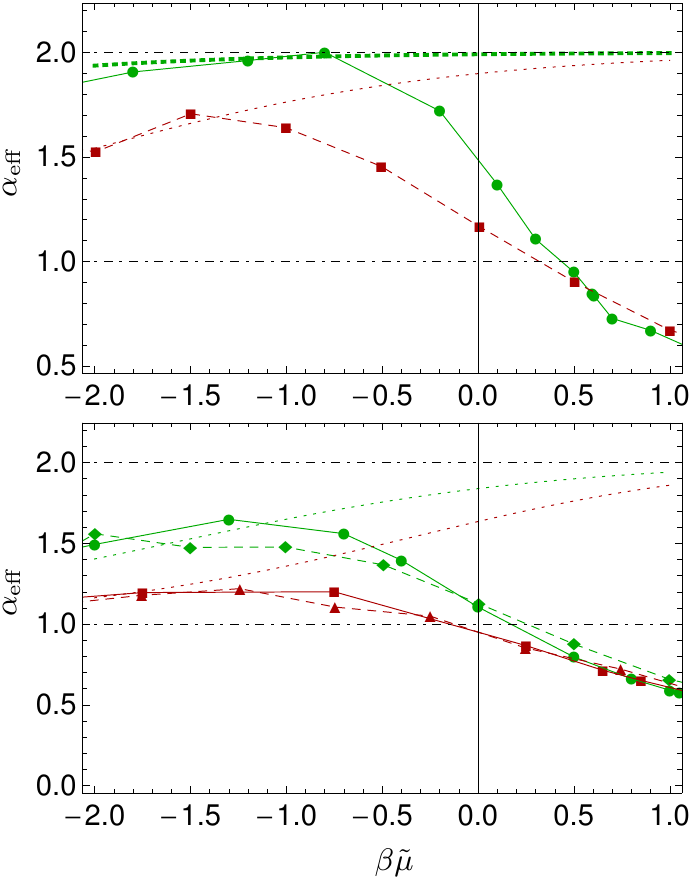}
\caption{The effective temperature exponent $\alpha_{\rm eff}$ defined in Eq.\ (\ref{temp4}) indicates the deviation of the system from a bosonic and fermionic Boltzmann gas ($\alpha_{\rm eff}=2$ and $\alpha_{\rm eff}=1$), shown as dotdashed lines in both panels, for $\beta\tilde{\mu}\in\mathcal{I}=[-2,0]$. Furthermore, $\alpha_{\rm eff}$ has been obtained from a numerical derivative of the data for the PSD from LW theory (\cite{PhysRevLett.112.135302}) and QMC (\cite{PhysRevLett.115.115301}). This is not meant to be a precise characterization of the corresponding EOS, but should rather indicate the overall trend. \emph{Upper panel.} The upper panel shows the predictions for $\beta\vare_{\rm B}=5$ from LW theory (green solid line) and for $\beta \vare_{\rm B}=3$ from QMC (red dashed line). The dotted lines are the corresponding curves from second order virial expansion. We observe $\alpha=2$ to be a good approximation of the upper curve in $\mathcal{I}$. \emph{Lower panel.} We compare LW (solid lines), QMC (dashed lines) and the virial expansion (dotted line) for $\beta\vare_{\rm B}=2$ (upper curves) and $\beta\vare_{\rm B}=0.5$ (lower curves). The data suggest to interpolate between the virial curves and $\alpha=1$ in the fitting range $\mathcal{I}$.}
\label{FigAlpha}
\end{figure}

In Fig. \ref{FigAlpha} we show $\alpha_{\rm eff}$ obtained from a numerical derivative of the LW and QMC data for the PSD. We observe that $\alpha_{\rm eff}$ deviates from the second order virial expansion in the interval $\mathcal{I}$ for small values of $\beta\vare_{\rm B}$, where it is close the fermionic Boltzmann limit $\alpha=1$. For $\beta\vare_{\rm B}\gtrsim 5$, the bosonic Boltzmann approximation $\alpha_{\rm eff}\approx 2$ is still a good account of the low-density region. The deviation from the second order virial expansion formula becomes relevant for $\beta\vare_{\rm B}\approx 2$, where $\alpha_{\rm eff}$ interpolates between the second order virial expansion value and 1 in the interval $\mathcal{I}$.

From this analysis we  get a better understanding of the transition from a Bose to a Fermi gas. Indeed, when applying a Boltzmann formula to the  low-density region of the density distribution, we need to specify whether the degrees of freedom of the gas are bosonic or fermionic, i.e., whether the particles that behave classically are dimers or atoms. From the proximity of $\alpha_{\rm eff}$ to 2 for $\beta\vare_{\rm B}\gtrsim 5$ we conclude that a good definition of the Boltzmann formula for our parameters is given by
\begin{align}
 \label{temp5} \text{B:}\ n_\sigma  &= \frac{\alpha}{\lambda_T^2} e^{\alpha\beta\tilde{\mu}},\ \alpha=\begin{cases} 2 & (B\leq 832),\\ 1 & (B> 832).\end{cases}
\end{align}
The discontinuity in this definition leads to a jump in temperature obtained from the Boltzmann fit, see Tab. \ref{TabSomTemps}.
Note that we choose $\tilde{\mu}$ for the fermions instead of $\mu$  in Eq. (\ref{temp5}). Hence we have $n_\sigma \lambda_T^2 \propto \tilde{z}^\alpha$ in the Boltzmann limit with $\tilde{z}=e^{\beta\tilde{\mu}}$. This finding is motivated by the values obtained with the fit as summarized in Tab. \ref{TabSomChems}, but does not affect our data analysis which is independent of the Boltzmann chemical potential.

\begin{table}[t!]
\begin{tabular}{|c||c|c|c|c|c|}
\hline $B[\text{G}]$ & $T_{\rm TOF}$ & $T_{\rm HF}$ & $T_{\rm B}$ & $T_{\rm V}$ & $T$  \\
\hline 692 & 55(18) & 61(8) & $\alpha=2$:\ 51(4) & 51(4) & 61(8)\\
\hline 732 & 55(20) & 52(5) & $\alpha=2$:\ 49(5) & 49(5) & 49(5)\\
\hline 782 & 46(14) & 36(3) & $\alpha=2$:\ 47(5) & 47(5) & 47(5)\\
\hline 812 & 57(9) & - & $\alpha=2$:\ 45(7) & 45(7) & 45(7)\\
\hline 832 & 59(13) & -& $\alpha=2$:\ 40(8) & 37(6) & 39(7)\\
\hline 852 & 58(14) & - & $\alpha=1$:\ 22(4) & 36(5) & 29(4)\\
\hline 892 & 71(19) & - & $\alpha=1$:\ 23(3) & 33(3) & 28(3)\\
\hline 922 & 67(19) & - & $\alpha=1$:\ 23(4) & 30(5) & 27(5)\\
\hline 952 & 57(11) & - & $\alpha=1$:\ 20(3) & 26(4) & 23(3)\\
\hline 982 & 67(26) & - & $\alpha=1$:\ 23(5) & 28(5) & 26(5)\\
\hline 1042 & 45(15) & - & $\alpha=1$:\ 23(6) & 26(6) & 25(6)\\
\hline
\end{tabular}
\caption{Lowest temperatures (in nK) attained in our experiment. We compare the results from fitting with different reference EOSs. The first column corresponds to the temperature obtained from a Boltzmann fit of the tail of the momentum distribution after time-of-flight (TOF) \cite{PhysRevLett.114.230401}. The corresponding values are the mean and standard deviation after averaging over $\sim30$ shots. The remaining columns show the median and median deviation of fitting the low density region of the in-situ density profile to the Hartree--Fock temperature $T_{\rm HF}$ (\ref{temp1}), Boltzmann temperature $T_{\rm B}$ with $\alpha$ (\ref{temp5}), and second order virial expansion $T_{\rm V}$ (\ref{temp2}). The last column displays our best estimate for $T$ as discussed in the text. Note that we determine the temperature for each shot individually, such that the mean values and their error displayed here only serve as an orientation.}
\label{TabSomTemps}
\end{table}

\begin{table}[t!]
\begin{tabular}{|c||c|c|c|c|c|}
\hline $B[\text{G}]$ & $\tilde{\mu}_{0,\rm TF}$ & $\tilde{\mu}_{0,\rm HF}$ & $\tilde{\mu}_{0,\rm B}$ & $\mu_{0,\rm V}+\frac{\vare_{\rm B}}{2}$ & $\mu_{0,\rm V}$\\
\hline 692 & 34(3) & 39(3) & 49(5) & - & -\\
\hline 732 & 51(4) & 60(5) & 65(5) & - & -\\
\hline 782 & 78(6) & 91(6) & 75(8) & - & -\\
\hline 812 & 110(7) & - & 100(6) & 110(10) & -180(10)\\
\hline 832 & 130(7) & - & 120(10) & 120(6) & 40(6)\\
\hline 852 & 160(8) & - & 170(6) & 140(8) & 110(8)\\
\hline 892 & 200(7) & - & 200(7) & 180(9) & 170(9)\\
\hline 922 & 210(10) & - & 200(10) & 190(10) & 190(10)\\
\hline 952 & 190(10) & - & 180(10) & 170(10) & 170(10)\\
\hline 982 & 210(10) & - & 200(7) & 190(6) & 190(6)\\
\hline 1042 & 200(10) & - & 180(20) & 180(10) & 180(10)\\
\hline
\end{tabular}
\caption{Central chemical potentials in nK. The first column shows the mean and standard deviation for the Thomas--Fermi (TF) value $\tilde{\mu}_{0,\rm TF}$ which is used in the main text. From the HF-, B-, and V-fits of Tab. \ref{TabSomTemps} we obtain effective chemical potentials, which are close to the TF result. We show their median and median deviation values. The 2D binding energy $\vare_{\rm B}$ which relates $\tilde{\mu}_0$ and $\mu_0$ is given in Tab. \ref{Taba2D} below.}
\label{TabSomChems}
\end{table}

Note that $\alpha=1$ is \emph{a priori} not a lower bound, and this is also not suggested by the theoretical data for even smaller $\beta\vare_{\rm B}$. Still, for the interaction parameters $5\gtrsim\beta\vare_{\rm B}\gtrsim 0.5$ relevant to Fig. 3 of the main text, the best available theoretical predictions indicate that $\alpha_{\rm eff}$ interpolates between the second order virial result and the fermion Boltzmann formula $\alpha=1$ in the interval $\mathcal{I}$. 

Denoting the temperature obtained from a fit of Eq. (\ref{temp2}) to the low-density region by $T_{\rm V}$ and the one using Eq. (\ref{temp5}) by $T_{\rm B}$, we thus conclude that
\begin{align}
 \label{temp6} T= \frac{T_{\rm V}+T_{\rm B}}{2}
\end{align}
is a good estimate for the temperature of the sample for $\beta\vare_{\rm B}\gtrsim 0.5$: On the Bose side, $T_{\rm V}\simeq T_{\rm B}$ and the average is trivial. On the Fermi side, our analysis of $\alpha_{\rm eff}$ indicates that $T_{\rm V}$ and $T_{\rm B}$ are upper and lower bounds to the actual temperature. We apply the temperature definition in Eq. (\ref{temp6}) for the magnetic fields $732\leq B\text{[G]}\leq 1042$. The systematic error for $B[\text{G}]>892$ where $\alpha_{\rm eff}$ becomes smaller than 1 is not relevant for the results on the EOS presented in our article. Only for $B=692\text{G}$ we apply the better estimate from HF theory as a Boltzmann fit is clearly disfavoured by the tail of the density profile.

In Tab. \ref{TabSomTemps} we summarize the results of the temperature fit using the reference formulas presented in this section. We indeed find the bosonic Boltzmann values to be close to the result from the virial expansion for $B\leq 832\text{G}$. For larger magnetic fields we confirm the expectation that the virial fit is consistently above the fermionic Boltzmann result, although both approach each other. For $B\geq 892\text{G}$ the deviation between $T_{\rm V}$ and $T_{\rm B}$ is 10nK or less. We note that $T_{\rm V}$ gives a smooth temperature variation as we cross the Feshbach resonance from low to high magnetic fields. In contrast, $T_{\rm B}$ jumps due to the discontinuous definition of $\alpha$ in Eq. (\ref{temp5}). We conclude that the smooth phenomenological interpolation for $\alpha$ employed in \cite{PhysRevLett.114.230401} is justified by the second order virial expansion. 

Further note that the decrease in temperature from the Bose to the Fermi side of the crossover is approximately by a factor of two. This can be explained by assuming an adiabatic ramp across the Feshbach resonance when changing the interaction strength. Since the degrees of freedom double at constant entropy, the temperature is reduced by a factor of two. We find a good agreement with the temperature obtained from the large momentum tail of the momentum distribution after TOF on the Bose side, but find substantial differences on the Fermi side, as  has already been discussed in Ref.\ \cite{PhysRevLett.114.230401}. For the present analysis we obtain the temperature from fits of the in-situ density profiles as this gives an overall consistent picture.

From the fit of the outer regions of the cloud we obtain, as a by-product, an estimate for the central chemical potential $\mu_0$. We summarize the corresponding values in Tab. \ref{TabSomChems}. We also show the chemical potential determined by the TF fit of the high-density central region. We find good agreement within the errors from the different methods. This indicates that the reference EOSs introduced in this section are well-applicable to the wings. Furthermore, deviations in $\mu_0$ obtained from approximate reference EOSs does not affect our data analysis based on $\tilde{\mu}_{0,\rm TF}$, which we have shown to be robust.

\section{2D Scattering Physics}
\label{App2Dscat}
Due to the tight confinement along the $z$-direction, the scattering physics of the 3D atomic gas can be mapped to those of a 2D gas. Here we summarize the formulas which are required to translate the 3D scattering length $a_{\rm 3D}(B)$ across the Feshbach resonance at $832.2\text{G}$ \cite{PhysRevLett.110.135301} to the 2D scattering length characterizing the quasi-2D system. For a more detailed discussion of similar aspects see also Refs.\ \cite{PhysRevA.64.012706,PhysRevLett.112.045301,PhysRevLett.114.230401}.

Scattering in 2D is peculiar in the sense that the scale dependence of interaction corrections is typically logarithmic. This is reflected in the universal low-momentum behavior of the 2D scattering amplitude given by
\begin{align}
 f_{\rm 2D}(k) = \frac{4\pi}{-\ln(k^2a_{\rm 2D}^2)+\rmi \pi}
\end{align}
for $k\to0$, where $k$ is the relative momentum of two colliding particles in the center-of-mass frame. Interaction corrections are important if $f(k_0)$ becomes of order unity, where $k_0$ is a typical momentum of scattering particles.

The scattering amplitude for the 3D system which is strongly harmonically confined in the $z$-direction with an oscillator length $\ell_z$ is given by \cite{PhysRevA.64.012706,PhysRevLett.112.045301}
\begin{align}
 \label{scat2} f_{\rm q2D}(k) = \frac{4\pi}{\sqrt{2\pi}\frac{\ell_z}{a_{\rm 3D}}+w(k^2\ell_z^2/2)},
\end{align}
where the function $w(\xi)$ reads
\begin{align}
 \nonumber w(\xi) &=\lim_{J\to \infty}\Bigl[\sqrt{\frac{4J}{\pi}}\ln\Bigl(\frac{J}{e^2}\Bigr)\\
 \label{scat3} &-\sum_{j=0}^J \frac{(2j-1)!!}{(2j)!!} \ln(j-\xi-\rmi 0)\Bigr].
\end{align}
For low density $\xi\ll1$ we have $w(\xi)\simeq w_{\rm lim}(\xi) = -\ln(2\pi\xi/A)+\rmi \pi$ with $A=0.905$. We define the filling correction
\begin{align}
 \label{scat4} \Delta w(\xi) = w(\xi) - w_{\rm lim}(\xi)\geq0.
\end{align}
For $\xi<0$ we define $\Delta w(\xi)=0$. If $\ell_z$ is sufficiently small (tight confinement), we have $\Delta w\approx0$, and $f_{\rm q2D}(k)$ approaches the form of $f_{\rm 2D}(k)$  with $a_{\rm 2D}$ given by
\begin{align}
 \label{scat5} a_{\rm 2D}^{(0)} = \ell_z \sqrt{\frac{\pi}{A}} \exp\{-\sqrt{\frac{\pi}{2}}\frac{\ell_z}{a_{\rm 3D}}\}.
\end{align}

Neglecting $\Delta w$ is justified if the relevant momentum scale of scattering, $k_0$, satisfies $k_0\ell_z\ll 1$. In a many-body setting, the corresponding momentum scale is given by $\hbar k_0=\sqrt{2M\tilde{\mu}}$ \cite{PhysRevLett.112.045301}, thus $k_0^2\ell_z^2/2=\tilde{\mu}/\hbar\omega_z$. If $k_0\ell_z$ is not small, the effective 2D scattering length of the many-body system receives a $k_0\ell_z$-dependent correction. The latter is obtained from keeping the higher order terms in $\xi$ in the denominator of Eq. (\ref{scat2}) when equating $f_{\rm 2D}(k_0)=f_{\rm q2D}(k_0)$. We then arrive at the more general formula
\begin{align}
 \label{scat6} a_{\rm 2D} = \ell_z \sqrt{\frac{\pi}{A}} \exp\{-\sqrt{\frac{\pi}{2}}\frac{\ell_z}{a_{\rm 3D}}\} e^{-\frac{1}{2}\Delta w(k_0^2\ell_z^2/2)}.
\end{align}
For the lowest temperatures attained in the experiment, $\xi$ is irrelevant in the bosonic limit, but reaches $\xi\simeq 0.7-0.9$ for the highest magnetic fields on the Fermi side. The shift of $\ln(k_{\rm F}a_{\rm 2D})$ with respect to $\ln(k_{\rm F}a_{\rm 2D}^{(0)})$, however, is only $-\frac{1}{2}\Delta w = -(0.4-0.5)$ in this case.

We define the 2D binding energy as
\begin{align}
\vare_{\rm B} = \frac{\hbar^2}{Ma_{\rm 2D}^2}.
\end{align}
Analogously, with $\vare_{\rm B}^{(0)}=\hbar^2/(M[a_{\rm 2D}^{(0)}]^2)$ we have $\vare_{\rm B}=\vare_{\rm B}^{(0)}e^{\Delta w}$. The effect of $\Delta w$ appears now exponentiated. However, the dependence of the PSD on $\beta\vare_{\rm B}$ is only logarithmic. Note that $\vare_{\rm B}$ does in general not coincide with the energy $\vare_{\rm B}^{\rm C}$ of the confinement induced bound state \cite{RevModPhys.80.885} given by the solution of $\ell_z/a_{\rm 3D}=f_1(\vare_{\rm B}^{\rm C}/(\hbar\omega_z))$ with
\begin{align}
 f_1(\Omega) = \int_0^\infty \frac{\mbox{d}u}{\sqrt{4\pi u^3}} \Bigl(1-\frac{e^{-\Omega u}}{\sqrt{(1-e^{-2u})/(2u)}}\Bigr).
\end{align}
We compare the values of $\vare_{\rm B}$, $\vare_{\rm B}^{(0)}$, and $\vare_{\rm B}^{\rm C}$ for all magnetic fields $692-1042\text{G}$ in Tab.\ \ref{Taba2D}.

\section{Systematic uncertainties}

In the main text we display the statistical errors of our determination of the EOS. Due to systematic uncertainties in the measurement of the density profiles $n(\vec{r})$, from which the EOS is constructed with the LDA, we have further systematic errors which are summarized here. Note that these systematic errors do not cover the deviations which arise from our choices of the fit function to determine the thermodynamic parameters $T$ and $\mu_0$. The latter can hardly be quantified at this point. The systematic errors of the experiment have already been discussed in detail in Ref.\ \cite{PhysRevLett.114.230401}.

\begin{table*}[t]
\begin{tabular}{|c|c|c|c|c|c|c|c|c|c|c|c|}
\hline $B\ \text[G]$  & $a_{\rm 2D}^{(0)}[\mu\text{m}]$  & $a_{\rm 2D}[\mu\text{m}]$ & $\vare_{\rm B}^{(0)}[\text{nK}]$ & $\vare_{\rm B}[\text{nK}]$ & $\vare_{\rm B}^{\rm C}[\text{nK}]$ & $\tilde{\mu}_{0}[\text{nK}]$ & $\vare_{\rm F}[\text{nK}]$ & $\ln(k_{\rm F}a_{\rm 2D})$ & $\tilde{\mu}_0/\vare_{\rm F}$ & $1/c$\\
\hline
\hline 692 & 0.000137 & 0.000128 & 4.31 $\times 10^9$ & 4.96 $\times 10^9$ & 13600. & 33(4) & 1180(123)& -7.27(5)$\binom{+8}{-6}$ & 0.028(4)$\binom{+4}{-4}$ & 0.024(2)$\binom{+4}{-4}$\\
\hline 732 & 0.00703 & 0.00624 & 1.63 $\times 10^6$ & 2.07 $\times 10^6$ & 4330. & 50(6) & 951(132) & -3.49(7)$\binom{+8}{-6}$ & 0.052(10)$\binom{+8}{-8}$ & 0.046(4)$\binom{+7}{-7}$\\
\hline 782 & 0.147 & 0.12 & 3730. & 5610. & 766. & 78(7) & 590(44) & -0.76(3)$\binom{+8}{-6}$ & 0.13(2)$\binom{+2}{-2}$ & 0.12(1)$\binom{+2}{-2}$\\
\hline 812 & 0.517 & 0.376 & 302. & 569. & 191. & 113(9) & 469(36) & 0.29(3)$\binom{+6}{-5}$ & 0.24(3)$\binom{+3}{-3}$ & 0.21(2)$\binom{+3}{-3}$\\
\hline 832 & 1.02 & 0.699 & 77.3 & 165. & 65.5 & 130(10) & 417(45) & 0.86(5)$\binom{+6}{-5}$ & 0.31(4)$\binom{+4}{-4}$ & 0.28(3)$\binom{+4}{-3}$\\
\hline 852 & 1.83 & 1.15 & 24. & 61.5 & 22.6 & 154(13) & 371(41) & 1.31(5)$\binom{+6}{-5}$ & 0.41(6)$\binom{+6}{-5}$ & 0.38(4)$\binom{+5}{-5}$\\
\hline 892 & 4.73 & 2.45 & 3.61 & 13.5 & 3.58 & 193(16) & 366(35) & 2.10(8)$\binom{+6}{-5}$ & 0.53(7)$\binom{+7}{-7}$ & 0.48(5)$\binom{+7}{-6}$\\
\hline 922 & 8.29 & 4.01 & 1.17 & 5.00 & 1.17 & 204(18) & 341(40) & 2.58(8)$\binom{+6}{-5}$ & 0.60(9)$\binom{+8}{-7}$ & 0.54(6)$\binom{+7}{-7}$\\
\hline 952 & 13.2 & 6.96 & 0.46 & 1.66 & 0.459 & 191(19) & 295(37) & 3.02(7)$\binom{+6}{-5}$ & 0.65(10)$\binom{+9}{-8}$ & 0.59(6)$\binom{+8}{-7}$\\
\hline 982 & 19.6 & 9.48 & 0.209 & 0.897 & 0.209 & 205(14) & 319(25) & 3.40(8)$\binom{+6}{-5}$ & 0.64(7)$\binom{+9}{-8}$ & 0.59(6)$\binom{+8}{-7}$\\
\hline 1042 & 36.5 & 18.4 & 0.0605 & 0.237 & 0.0605 & 197(21) & 268(33) & 3.96(10)$\binom{+6}{-5}$ & 0.74(12)$\binom{+10}{-9}$ & 0.67(8)$\binom{+9}{-8}$\\
\hline
\end{tabular} 
\caption{Scattering parameters and low temperature EOS as shown in Fig. 1. We show the result of averaging the observables over approximately 30 shots for each magnetic field. The error is shown as $(\text{stat.})\binom{+\rm{sys.}}{-\rm{sys.}}$, where the statistical error is given by the standard deviation and the systematic error results from the systematic uncertainties discussed in this section. Note that the 2D binding energy $\vare_{\rm B}$ is generally larger than the quasi-2D universal dimer energy $\vare_{\rm B}^{\rm C}$.}
\label{Taba2D}
\end{table*}

\textbf{Absorption imaging.} The planar density profile $n(\vec{r})$ is constructed from the (column integrated) optical density $\text{OD}(\vec{r})$ after absorption imaging along the $z$-direction. At zero detuning we have
\begin{align}
n(\vec{r})\sigma_0^* = \text{OD}(\vec{r}) + \frac{I_0(\vec{r})}{I_{\rm sat}^*}(1-e^{-\text{OD}(\vec{r})}),
\end{align}
where $\sigma_0^*$ is the effective absorption cross section, $I_0$ the initial intensity of the probe beam before the atomic cloud, and $I_{\rm sat}^*$ is the effective saturation density. The experiments are performed for $I_0/I_{\rm sat}^*=0.97^{+0.13}_{-0.08}$. This leads to a systematic error in the peak density $n_{\rm peak}\mbox{}^{+7\%}_{-4\%}$. Furthermore, due to the nonzero binding energy, the optical transition frequencies of dimers are shifted, which leads to a reduced absorption cross section $\sigma_0^*$ on the Bose side. We compensate for this effect for $B[\text{G}]=692, 732, 782$ by means of a rescaling in $\sigma_0^*$. The systematic uncertainty in this rescaling factor leads to an additional systematic error in the peak density $n_{\rm peak}$ of at most $8\%$.

\textbf{Atoms in noncentral pancakes.} Our experiment realizes the planar gas in a pancake shaped geometry at $z=0$. Therein the pancake is realized as a standing-wave optical dipole trap from interfering two laser beams at a small angle each. In this way, further pancakes at $z\gtrless0$ are created at a distance $\simeq 4\mu\text{m}$ from the central pancake. A nonvanishing population of the noncentral pancakes leads to an overestimation of the planar density. Assuming the gas in the higher pancakes to be a thermal gas we find  that the density is overestimated at most by $10\%$.

\textbf{Trap parameters.} The application of the LDA requires an accurate knowledge of the in-plane trapping potential $V(\vec{r})$. For the purpose of the present analysis we assume
\begin{align}
 V(\vec{r}) = \frac{M}{2} ( \omega_x^2x^2+\omega_y^2y^2) + c_{4x} x^4 + c_{4y}y^4 + c_{4xy} x^2y^2,
\end{align}
where the potential is centered at $\vec{r}_0=0$, and $\omega_x = \omega_y$ \cite{PhysRevLett.114.230401}.The harmonic trapping frequencies $\omega_{x/y}$ are known to an accuracy of $0.4\%$, hence their uncertainty can safely be neglected in the following. The quartic terms $c_4$ are only known to a precision of $25\%$, but they are small and hardly contribute to the density region of interested. Higher order terms in the expansion are not relevant as the gas does not extend to regions so far from the center. 

\textbf{Magnification.} The magnification of the imaging system relates the pixels of the imaged density profile to $\vec{r}[\mu\text{m}]$ in $n(\vec{r})$. The magnification factor is known to an uncertainty of $3\%$, which results in an error of $6\%$ in $V(\vec{r})$, where the distance appears squared.

\textbf{Binding energy.} The determination of the binding energy from $\vare_{\rm B}=\vare_{\rm B}^{(0)}e^{\Delta w(\tilde{\mu}_0/(\hbar\omega_z))}$ requires knowledge of $a_{\rm 3D}(B)$ and $\ell_z$ to compute $\vare_{\rm B}^{(0)}$. Both quantities, however, are known very well in our experiment and their error can be neglected  \cite{PhysRevLett.114.230401}. On the other hand, for large $\tilde{\mu}_0$, the deviation due to the exponential of $\Delta w$ can be substantial. We have $\hbar \omega_z/k_{\rm B} = 265 \text{nK}$. Assuming a deviation of $\tilde{\mu}_0$ by $\pm 10\%$ from the mean values $\tilde{\mu}_{0,\rm TF}$ displayed in Tab. \ref{TabSomChems} we find the systematic error in $\vare_{\rm B}$ to be $\pm 4\%$.

The listed systematic uncertainties affect the thermodynamic variables $n, \mu$, and $T$. The density is influenced by the uncertainties in the absorption imaging. This leads to an uncertainty $n\mbox{}^{+16\%}_{-13\%}$ for $B\leq 782\text{G}$ (Bose side) and $n\mbox{}^{+12\%}_{-11\%}$ for $B\geq 812\text{G}$ (crossover regime and Fermi side). The chemical potential is derived from the LDA assumption $\mu(\vec{r})=\mu_0-V(\vec{r})$ and hence inherits the systematic uncertainties in $V(\vec{r})$. Those result from the uncertainty in the trapping frequencies and the magnification of the imaging system, giving $\mu^{+6\%}_{-6\%}$. The temperature is fitted in the low-density region with a reference EOS of the form $n_\sigma = \lambda_T^{-2} \nu(\beta\mu_0-\beta V(\vec{r}))$, where $\nu(x)$ is a dimensionless function of $x=\beta\mu$. The uncertainties of the density influence both $\lambda_T^{-2}$ and $\mu_0$. For instance, in the Boltzmann case we have $n_\sigma = \frac{e^{\beta\mu_0}}{\lambda_T^2} e^{-\beta V(\vec{r})} = C e^{-\beta V(\vec{r})}$, such that density uncertainties are absorbed in the irrelevant prefactor $C$. We conclude that the systematic uncertainty in $T$ is the same as the one for $\mu$, i.e. $\pm 6\%$.

\begin{figure}[ht!]
\centering
\includegraphics[width=8.5cm]{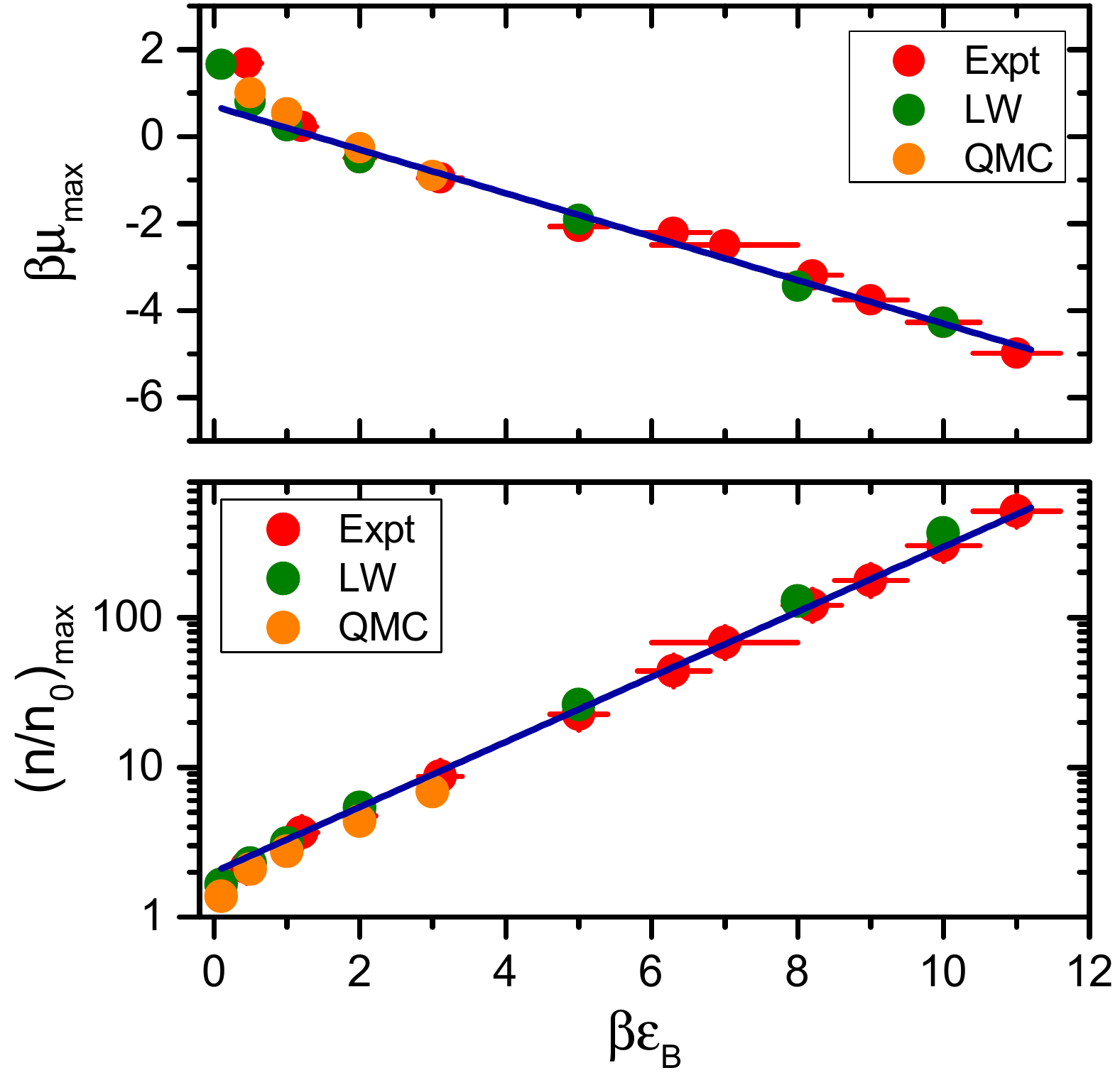}
\caption{Comparison of the measured location and height of the maximum of the curve $n/n_0$ (red, ``Expt'') with theoretical predictions from LW (\cite{PhysRevLett.112.135302}, green) and QMC (\cite{PhysRevLett.115.115301}, orange) theory. In the upper panel we show the location of the maximum $(\beta\mu)_{\rm max}$ for fixed $\beta\vare_{\rm B}$. In addition to the expected linear scaling with $-\beta\vare_{\rm B}/2$ we find a subleading correction of $\approx \ln(2)$. The solid blue line is the curve $(\beta \mu)_{\rm max}=-\beta\vare_{\rm B}/2+\ln(2)$. In the lower panel we show the height $(n/n_0)_{\rm max}$ of the maximum as a function of $\beta \vare_{\rm B}$. The blue solid curves is the reference $(n/n_0)_{\rm max}=2e^{\beta \vare_{\rm B}/2}$.}
\label{FigMax}
\end{figure}

\section{Maximum of the equation of state}
The EOS at nonzero temperature expressed by the function $n/n_0=h(\beta\mu,\beta\vare_{\rm B})$ exhibits a maximum as a function of $\beta \mu$ for fixed $\beta \vare_{\rm B}$. Here we quantify the location and height of this maximum from our experimental data and compare to theoretical predictions from QMC and LW calculations.

Our data is consistent with a maximum of height $(n/n_0)_{\rm max}\simeq2 e^{\beta\vare_{\rm B}/2}$ at $(\beta \mu)_{\rm max}\simeq-\frac{\beta \vare_{\rm B}}{2}+\ln(2)$. Writing $x=\beta\mu$ and $y=(n/n_0)$, the maxima thus all approximately lie on the curve $y=4 e^{-x}$. We demonstrate this behavior in Fig. \ref{FigMax}, where we also compare to theory.

\vfill

\end{document}